# A FORMAL APPROACH TO FIREWALLS TESTING TECHNIQUES

Alexander Barabanov (NPO «Echelon», 107023, Moscow, Russia, Elektrozavodskaya st., 24, mail@npo-echelon.ru)

Alexey Markov (Bauman's Moscow State Technical University, 107005, Moscow, Russia, 2nd Baumanskaya st.,5, mail@cnpo.ru)

Valentin Tsirlov (Bauman's Moscow State Technical University, 107005, Moscow, Russia, 2nd Baumanskaya st.,5, vz@cnpo.ru)

Traditional technologies of firewall testing are overlooked. A new formalized approach is presented. Recommendations on optimization of test procedures are given.

**Introduction**

Basic means for creation of any network protection system are firewalls (FW), the number of various families of such devices are counting up to several hundreds. Existing typical techniques of testing FW have descriptive character and this makes difficult the automation and optimization of evaluation processes for matching computer networks and means of network protection. The article describes the approach to formalization of FW testing techniques which allow to identify the factors connected with time, cost and completion of FW test.

**Formalized description of technique for carrying out tests**

Let $\Sigma$ be a firewall - local (one component type) or functionally distributed software (software-hardware) tool (complex) which executes the information filtering, for information entering the automated system (AS) and/or leaving the AS.

Let $R = \{r_i\}$ – multiple demanded requirements for FW $\Sigma$, $T = \{t_i\}$ – multiple test procedures which check execution of demanded requirements.

The technique of test procedure development is displayed by the following: $M: \Sigma \times R \rightarrow T$. Function M is based on the requirement $r_i \in R$ and information on FW execution is $\Sigma$. Note, that M function for the given ID $\Sigma$ is considered to be bijective display:

- $\forall r_1, r_2 \in R$ is carried out $(r_1 \neq r_2) \Rightarrow (M(\Sigma, r_1) \neq M(\Sigma, r_2))$, so that different security requirements take part in formation of various text procedures;
- $\forall t_i \in T \; \exists \; r_i \in R: M(\Sigma, r_i) = t_i$ – any test procedure is developed with an aim to check a certain requirement.

Each test procedure $t_i \in T$ is defined by the execution objective, order of executed actions, result of the test procedure execution and criteria of acceptance of favorable solution [1,2].

*The test objective* contains statement of intention about the execution of FW conformance evaluation according to the demanded requirements. *The order of executed actions* defines the sequence of steps which are carried out by the expert for restoration of FW back to the normal state and carrying out the test procedure. *The execution results of test procedures* are recorded with the use of various software tools (ST) for test performance, for example: tools for generation and interception of network traffic, search of residual information, access delimitation system checks. *Acceptance criteria of favorable solution* must contain model results of test procedure execution. Enter the commands of requirements execution $F_R$ and validity of test procedure execution $F_C$.

Execution operator of requirement $r_i$ for FW $F_R: \Sigma \times R \rightarrow \{0,1\}$:

$$F_R(\Sigma, r_i) = \begin{cases} 1, \text{ if requirement } r_i \text{ is met}; \\ 0, \text{ if not.} \end{cases}$$

Operator of test procedure execution validity $t_i$ for FW $\Sigma$ $F_C: \Sigma \times T \rightarrow \{0,1\}$:

$$F_C(\Sigma, t_i) = \begin{cases} 1, \text{ if test } t_i \text{ is succesfully performed}; \\ 0, \text{ if not.} \end{cases}$$

Statement $F_C$ shows that execution of test procedure for FW $\Sigma$ was carried out successfully: Actual results which were registered during the test execution match the model values which are stated in the test procedure description.

The set of 5 objects $\mathbb{A} = \{\Sigma, R, M, F_R, F_C\}$ will be called a testing technique, where $R$ is a range of requirements, demanded from FW $\Sigma$, $M$ is a method of test procedure development, $F_R$ and $F_C$ are requirement execution and validity of the test procedure execution operators accordingly, also $F_R(\Sigma, r_i) \Rightarrow F_C(\Sigma, M(\Sigma, r_i))$ is valid for $\forall r_i \in R$

In overall view, the test consists of three stages: planning, testing and results analysis.

During *planning stage* the documentation and operating FW features analysis is carried out. Experts must determine that developer states that FW corresponds the R requirements in the technical documentation, in other words $F_R(\Sigma, r_i) = 1$ for $\forall r_i \in R$. Based on the documentation analysis, FW test runs and demanded requirements, the range of test procedures $T = \{t_i\}$ is formed, where $t_i = M(\Sigma, r_i)$.

*Testing* is carried out using a set of test procedures $T = \{t_i\}$, following that results of test procedure execution which have to be registered are identified for each test procedure.

On the analysis stage of the factual and sample values, set of sequenced pairs in form of $(t_i, F_C(\Sigma, t_i))$ are acquired. Requirement conformance is declared for FW $\Sigma$ compared to $R = \{r_i\}$, if:

$$\sum_{i=1}^{n} \left( F_R(\Sigma, r_i) \cdot F_C(\Sigma, M(\Sigma, r_i)) \right) = n,$$

in other words, during the test performance the compliance of factual capabilities of FW declared in the documentation or a normative document is determined [1,3].

**Technique of firewall tests concerning the compliance with information security requirements**

Requirements for FW concerning the information security may be defined as several security classes. Each class is defined by a minimum group of requirements. Let us examine the check order of most resource consuming requirements $R = \{r_1, r_2, r_3\}$ (examine table 1).

Table 1.

**Main requirements for internetwork screens**

| Name | Requirement |
|------|-------------|

| Name | Requirement |
|---|---|
| $r_1$ | FW has to provide filtration on network level. The filtration solution can be accepted for each network package independently (at least) from network addresses of sender and recipient or based on other equivalent attributes. |
| $r_2$ | FW has to provide identification and authentication of FW administrator on his request. FW must provide a capability of identification and authentification according to identifier (code) and password which is permanent by default settings. In addition FW must prevent access of unidentified subject or a subject who did not confirm his identity during the authentification. |
| $r_3$ | FW must contain means of control of its software and informational integrity. |

Typical scheme of testing stand constitutes of two network segments with computers which are separated by FW.

**Data filtration and address transmission checks**

The aim of performing a check is to identify the degree of conformity of FW functional capabilities concerning the network packages taking into account the following parameters: network address of the sender, network address of the recipient. Initial data for creating the test procedure $t_1$ are the multiple network addresses which are used in the test segments: $IP = IP_S \cup IP_R = \{IP_S^1, IP_S^2, \dots, IP_R^1, IP_R^2\}$. It is assumed that during test performance the packages are sent from the external network segment (network addresses in the form of $IP_S^i$) to the internal segment (network addresses in the form of $IP_R^j$).

The check consists of the following steps:

1. Adjustment of FW filter rules in accordance with checked requirement, as a result multiple denying $RULE^0 = \{rule_1^0, rule_2^0, \dots\}$ are being formed and allowing $RULE^1 = \{rule_1^1, rule_2^1, \dots\}$ rules of internetwork screening, moreover, each rule represents ordered set with the following structure: $rule_k^{0/1} = (IP_S^i, IP_R^j)$, where $IP_S^i$ is the network address of the sender, and $IP_R^j$ is the network address of the recipient.

2. Launch of interception and analysis of network packages ST in the internal and external network segments.

3. Generation of network packages from external segment to internal segment for different possibilities of pars: $packet_k = \left(IP_S^i, IP_R^j, payload^k\right)$.

4. Completion of network packages interception. As a result we receive the following sets of intercepted packages: $PACKET^{IN} = \{packet_1^{IN}, packet_2^{IN}, ...\}$ and $PACKET^{OUT} = \{packet_1^{OUT}, packet_2^{OUT}, ...\}$, where $packet_k^{IN} = \left(IP_S^i, IP_R^j, payload^k\right)$ – network packages which were intercepted in the external segment, $packet_k^{OUT} = \left(IP_S^i, IP_R^j, payload^k\right)$ – network packages intercepted in the inner segment.

5. Export of registration log for allowed and denied packages in FW. As an accomplishment result the set of entries are being created on denial of transmission $JOUR^0 = \{journal_1^0, journal_2^0, ...\}$ and allowance of transmission $JOUR^1 = \{journal_1^1, journal_2^1, ...\}$ of network package. Each entry is in the form of: $journal_k^{0/1} = \left(IP_S^i, IP_R^j\right)$.

Results of the test procedure performance are:

1. FW Configuration – sets $\{rule_i^0\}$ and $\{rule_i^1\}$.

2. Results of package interception in the outer and inner interfaces of FW– множества $\{packet_i^{IN}\}$ and $\{packet_i^{OUT}\}$.

3. Registration log fragment with FW events which demonstrates the results of network package filtering: sets $\{journal_i^0\}$ and $\{journal_i^1\}$.

Let us assume that we take conformity fixation of factual results (packages in the entry interface of FW, packages in the exit interface of FW and fragment of registration log for FW events) and expectable results (FW filter rules) as a criteria for acceptance of favorable solution:

$$\begin{cases} PACKET^{OUT} = RULE^1; \\ PACKET^{IN}/PACKET^{OUT} = RULE^0; \\ PACKET^{OUT} = JOUR^1; \\ PACKET^{IN}/PACKET^{OUT} = JOUR^0. \end{cases}$$

During the test performance following ST's can be used: *nmap, Packet Generator* (network packages generation), *wireshark, tcpdump* (interception and analysis of network packages), software complex *«VS-Scanner»* (generation, interception and analysis of network packages).

The similar FW requirements are demanded from filter mechanisms on other levels ISO/OSI levels. For the channel level the requirement looks the following way: "FW has to provide filter taking into account exit and entry network interface as a tool for authenticity check of network addresses". The checking technique is similar to the technique described above, but the channel level addresses (MAC-addresses) are used as filtration criteria. The requirement according to filtration is checked on the network level taking into account any significant fields of network packages. As a rule during the test performance the attention has to be concentrated on the filter mechanism taking into account following fields of the network level package: sender address, recipient address, protocol type of upper (traffic) level, time of package life (TTL).

**Checking of administrator identification and authentication mechanisms**

The aim of the check is to determine the conformity degree of functional capabilities of the FW according to FW administrator identification and authentication.

Let us consider that $A$ is an alphabet of password and identifiers for FW administrators. Administrator identifier is defined $id \in ID \subseteq A^*$, password - $pwd \in PWD \subseteq A^*$. Administrator profile is $adm_i \in ADM$ is defined by the following sequence $adm_i = (id_j, pwd_k)$.

Enter the validity operator for the account data $F_{AUT}: ADM \rightarrow \{0,1\}$:

$$F_{AUT}(adm) = \begin{cases} 1, \text{ recieved administrator access}; \\ 0, \text{ if not.} \end{cases}$$

It is assumed that identification/authentication is performed with the use of network protocols of computer in the internal network segment. Then the check will include the following sequence of actions:

1. Activation of identification and authentication FW mechanism and creation of multiple FW administrator account entries $ADM = \{adm_1, adm_2, ...\}$.

2. Launch of interception and network package analysis ST in the inner network segment.

3. Execution of identification and authentication conduction requests with the use of various account data combinations: Registered/unregistered identifier, valid/invalid password – $try_i = (id_j, pwd_k)$.

4. Network package generation from the internal network to the outer network (or vice versa) which are allowed/ denied to pass in accordance with the FW filter rules.

5. Completion of network package interception, export of registration log of FW events.

6. Analysis in order to find account data which is transferred in the open form.

During the execution of check for local identification/authentication accomplishment the steps 2 and 5 of the operational sequence are not performed.

Performance results for the test procedure are:

1. FW Configuration - multi *ADM* accounts for FW administrators.

2. Received the results of test requests for identification and authentication: multiple $\{F_{AUT}(try_i)\}$.

3. Results of network package interception in the internal and external FW interfaces.

4. Log fragment of FW events registration which demonstrates the results of identification and authentification.

Let us define the criteria for acceptance of favorable solution:

1. After the entry of registered identifier and password the user is granted access to the FW administrator tools: $F_{AUT}(try_i) = 1 \Leftrightarrow try_i \in ADM$.

2. After the entry of unregistered identifier and/or invalid password the user access to the FW administration tools is denied: $F_{AUT}(try_i) = 0 \Leftrightarrow try_i \notin ADM$.

3. The registration log includes all entries concerning the events of test attempts of access acquisition.

4. The Attempts of searching for identification data (username, password) in the intercepted packages did not provide any result.

The programs like *wireshark, tcpdump* of software complex "Scanner-VS" can be used for interception and network package analysis during the testing of identification/authentication mechanisms.

**Checking mechanisms of integrity control**

The check determines the conformity degree of FW functional capabilities concerning the integrity control of software and informational parts of the FW.

Let $FILE = \{file_1, file_2, \dots, file_n\}$ be the set of FW files (configuration files, program modules). Enter the operators of integrity violation $F_{MOD}$ and FW files integrity control $F_{INT}$.

Operator of integrity violation is $F_{MOD}: FILE \to \{0,1\}$:

$$F_{MOD}(file)$$
$$= \begin{cases} 1, \text{ file integrity is not violated during the test performance}; \\ 0, \text{ if not}. \end{cases}$$

The operator of FW files integrity control $F_{INT}: FILE \to \{0,1\}$:

$$F_{INT}(file) = \begin{cases} 1, \text{ File integrity is violated}; \\ 0, \text{ if not}. \end{cases}$$

Define $FILE^\Delta = \{file_1^\Delta, file_2^\Delta, \dots, file_n^\Delta\}$ – multiple FW files which were modified during the test conduction. Furthermore the modification of $file_i$ is performed in $file_i^\Delta$. During the validity check of FW integrity control mechanism execution the following action sequence can be used:

1. Activation of integrity control mechanism for software and informational FW part and identification of FW file set $FILE = \{file_1, file_2, \dots, file_n\}$.

2. Introduction of changes in the FW files (configuration alteration, substitution (modification) of executed files etc.) – acquirement of multiple modified files $FILE^\Delta = \{file_1^\Delta, file_2^\Delta, \dots, file_n^\Delta\}$.

3. Initialization of FW files integrity check (creating conditions which allow FW to perform integrity control).

4. Analysis of FW response to integrity violation of its software or informational part.

Results of test procedure accomplishment are assumed to be the following:

1. Sets of FW files: $FILE = \{file_1, file_2, \dots, file_n\}$.
2. Sets of modified FW files: $FILE^\Delta = \{file_1^\Delta, file_2^\Delta, \dots, file_n^\Delta\}$.
3. FW response to the integrity violation: $F_{INT}(file_1^\Delta), F_{INT}(file_2^\Delta), \dots, F_{INT}(file_n^\Delta)$.

Assuming we take the fact that all acts of integrity violation were located as a criteria for favorable solution acceptance:

$$F_{INT}(file_i^\Delta) = F_{MOD}(file_i).$$

**Recommendations on optimization of test performance procedure**

The optimization of FW test performance procedure task can be presented the following way. Let $\mathcal{T}: T \times \Sigma \to \mathbb{N}_0$ – time taken by the experts for conformance evaluation execution with the use of test procedure $t_i \in T$ for FW $\Sigma$. We define it with the help of $C: R \times \Sigma \to \mathbb{N}_0$ – the expenses for testing of requirement execution $R$ for FW $\Sigma$. Then the optimization procedure task for FW test performance is presented the following way (minimization of the FW test time along with expenses limitation):

$$\begin{cases} \sum_i \mathcal{T}(t_i, \Sigma) \to min, \\ \sum_i C(r_i, \Sigma) \leq C_\text{п}, \end{cases}$$

where $C_\text{п}$ – are limitations applied to expenses.

The most resource-consuming part of the tests is the testing of FW access control subsystem (network packages filter). The experience in FW testing helps to provide a number of recommendations which can reduce such expenses.

1. OS of test computers have to be loaded from removable storage (CD, USB storage device etc.) and must not require installation on the hard disc. This will allow to reduce the time expenses for installation and configuration of OS on test computers.

2. ST's which are required for test performance (for example, interception and network traffic analysis software tools or http-server) have to be included in the OS which is loaded on the test computers.

3. The program for centralized process generation and analysis of network packages has to be provided (program must be included in the OS which is loaded on test computers). The given program must have the filtering rule on entry which is checked during the particular check running and required information about the configuration of testing bench. Program must control the launch and deactivation of test ST (for generation and interception of network packages), to collect required information from test computers (for example, the list of intercepted packages) and to carry out analysis of the acquired results (comparison of factual results with the results set by the filter rules). Centralized control and information collection must be carried out in the computer networks which do not belong to internal or external segments. Aiming at minimization of financial expenses the control and information gathering can be carried out with the use of wireless technologies of data transfer.

4. The tools for the export of test results have to be provided.

The specified recommendations will allow decreasing the test time in the part of the following procedures execution:

- installation of OS and required ST on the bench computer;
- launch of interception and network package analysis ST in the internal network segment;
- launch of interception and network package analysis ST in the external network segment;
- generation of network packages from external network segment to the internal network segment;
- sign off of interception and network package analysis ST in the internal network segment;
- sign off of interception and network package analysis ST in the external network segment;
- creation of lists of accepted and denied network packages;
- comparative analysis of the accepted (denied) network packages and configured FW filter rules.

**Conclusion**

Proposed working approach for formalization of general approach and separate techniques allows to simplify test automation, FW and computer networks security verification in during the security execution.

Automation of the most routine steps (for example, synchronized launch of network traffic interception and generation of network packages ST) allows increasing the quality of performed test, reduce the time expenses and also reduce the number of possible test errors.

The approach can be recommended for carrying out various certification tests of information security products and attestation of informational support objects [3].